\documentclass[draftcls, onecolumn, 12pt]{IEEEtran}
\usepackage{amssymb,amsmath}
\usepackage[dvips]{graphicx}
\usepackage{amsfonts}
\usepackage{subfigure}
\usepackage{booktabs,multirow}
\usepackage{color}
\usepackage{makecell}
\usepackage{float}
\usepackage{colortbl}
\definecolor{mygray}{gray}{.7}

\IEEEoverridecommandlockouts

\begin{document}

\title{5G Wireless Backhaul Networks: Challenges and Research Advances}
\author{\normalsize
Xiaohu Ge$^1$,~\IEEEmembership{Senior~Member,~IEEE,} Hui Cheng$^1$, Mohsen Guizani$^2$,~\IEEEmembership{Fellow,~IEEE}, Tao Han$^1$,~\IEEEmembership{Member,~IEEE}\\
\vspace{0.70cm}
\small{
$^1$Department of Electronics and Information Engineering\\
Huazhong University of Science and Technology, Wuhan 430074, Hubei, P. R. China.\\
Email: \{xhge, hc\_cathy, hantao\}@mail.hust.edu.cn\\
\vspace{0.2cm}
$^2$Qatar University, Doha, Qatar\\
Email: mguizani@ieee.org}\\
\thanks{\small{ Submitted to IEEE Network SI on Unveiling 5G Wireless Networks: Emerging Research Advances, Prospects, and Challenges.}}
\thanks{\small{Corresponding author: Dr. Tao Han, Email: hantao@mail.hust.edu.cn }}
\thanks{\small{The authors would like to acknowledge the support from the International Science and Technology Cooperation Program of China (Grant No. 2014DFA11640 and 0903), the National Natural Science Foundation of China (NSFC) (Grant No. 61271224, 61301128 and 61471180), NFSC Major International Joint Research Project (Grant No. 61210002), the Hubei Provincial Science and Technology Department (Grant No. 2013BHE005), the Fundamental Research Funds for the Central Universities (Grant No. 2013ZZGH009 and 2014QN155), and EU FP7-PEOPLE-IRSES (Contract/Grant No. 247083, 318992 and 610524). }}
}

%\date{\today}
\renewcommand{\baselinestretch}{1.2}
\thispagestyle{empty}
\maketitle
\thispagestyle{empty}
%\newpage
\setcounter{page}{1}\begin{abstract}
 5G networks are expected to achieve Gigabit level throughput in future cellular networks. However, it is a great challenge to treat the 5G wireless backhaul traffic in an effective way. In this paper, we analyze the wireless backhaul traffic in two typical network architectures adopting small cell and millimeter wave communication technologies. Furthermore, the energy efficiency of wireless backhaul networks is compared for different network architectures and frequency bands. Numerical comparison results provide some guidelines for deploying future 5G wireless backhaul networks in economical and high energy efficiency ways.

\end{abstract}

%\begin{IEEEkeywords}
%Millimeter wave, 5G, user access, backhaul link.
%\end{IEEEkeywords}

\IEEEpeerreviewmaketitle

\newpage
\section{Introduction}

To meet the challenges of the expected traffic volume increase in wireless communications (say that of 2020 as compared to the 2010 level \cite{Taleb14}), research on the next generation cellular networks (or 5G networks), is highly anticipated in the next decade. Moreover, some potential transmission technologies are emerging to support thousand times wireless traffic volume increment in future wireless communications. The massive multiple-input multi-output (MIMO) antenna technology is validated to improve 10 to 20 times the spectrum efficiency in the same frequency bandwidth. The millimeter wave communication technology is explored to be applied to cellular networks, which can provide more than 100 MHz frequency bandwidths. Considering wireless signal propagation characteristics, the massive MIMO antenna and millimeter wave communication technologies will obviously reduce the cell coverage \cite{Bhushan14}. Therefore, small cell networks are emerging in 5G networks. In this case, 5G networks are not the simple upgrade of its predecessor, by adding additional spectrum and thus boosting the capacity, or replaced with advanced radio technology. It needs rethinking from the system and architecture levels down to the physical layer. In addition, we need to be able to answer the question of how to forward hundreds of Gigabit backhaul traffic in ultra dense cell networks with guaranteed quality of service (QoS) and affordable energy consumption by sustainable systems.

With the exponentially increasing demand for wireless data traffic in recent years, it is unfeasible for current cellular systems architecture to satisfy Gigabit level data traffic in an economical and ecological way \cite{Chen14}. One of the solutions is the small cell network, which is densely deployed by self-organizing, low-cost and low power small cell base stations (SBSs). In early studies, a low number of small cells is adopted to improve the signal-to-interference-and-noise ratio (SINR) of wireless links in limited hot areas, which is embedded in conventional cellular networks. In this case, a little burst backhaul traffic originating from small cells can be forwarded into the core network by traditional backhual link of cellular networks. When small cells are ultra densely deployed in cellular networks, it is a key problem to forward massive backhaul traffic into the core network. Moreover, there is a concern that the large number of small cells causes the signaling load on the networks nodes to increase due to frequent handovers and mobility robustness to be degraded due to increased handover failures and radio link failures \cite{Hoydis11}. The impact of small cell deployments on mobility performance in LTE-Advanced systems was investigated by system level simulations \cite{Yamamoto13}. Simulation results implied that the handover optimization technique can effectively decrease the handover failure rate. With rapidly developing in point-to-point microwave technologies, the wireless backhaul solution is becoming an attractive alternative for small cell networks. Based on simulation and measurement results, the microwave backhaul technology at high frequencies was a viable high-performance solution for wireless small cell backhaul links in non-line-of-sight (NLOS) \cite{Coldrey13}. Moreover, the high-performance NLOS backhaul link using higher frequencies compared to sub-5 GHz frequencies can provide the higher antenna gain for similar antenna sizes. This makes it possible to design small, compact, point-to-point fixed backhaul links with hundreds Gigabit per second throughput. 60 GHz and 70-80 GHz millimeter wave communication technologies for high capacity last mile and pre-aggregation backhaul were explored in \cite{Bojic13}. In addition, orthogonal frequency-division multiple (OFDM) access passive optical networks were discussed as the optical technology complement for enabling a flexible cost-efficient hybrid coverage. According to network simulation results for demanding urban small cell backhaul application, flexible high capacity hybrid millimeter wave/optical mobile backhaul networks presented a highly promising approach for future mobile backhaul networks. The coordinated multiple points (CoMP) technology is adopted in small cells to decrease the inter-site interference and improve the spectrum efficiency. However, the additional backhaul traffic is generated due to the possibility of sharing data among cooperative small cells. The backhaul bandwidth required by different coordination technologies was discussed in \cite{Samardzija09}. On the other hand, the energy efficiency of small cell networks is of great concern as the base station (BS) density will be significantly increased. Based on the random spatial network model, the energy efficiency of small cell networks was analyzed in \cite{Chang13}. Numerical results showed that the energy efficiency of small cell networks critically depends on the BS power consumption model.

Different with other studies in 5G networks, we focus on the throughput and energy efficiency of 5G wireless backhaul networks considering ultra dense small cells and millimeter wave communications. In detail, we first configure two typical small cell scenarios for comparison analysis. Then, we evaluate the wireless backhaul traffic models based on two typical small cell scenarios with different spectrum efficiencies. Furthermore, the energy efficiency of 5G wireless backhaul networks in two typical scenarios is analyzed by numerical results. Moreover, the impact of different frequency bands of wireless backhaul links on the energy efficiency of 5G backhaul networks is investigated for two typical scenarios. Finally, future challenges of 5G wireless backhaul networks are discussed and conclusions are drawn.

\section{System Model}

\begin{figure}[H]
%\begin{figure}
\begin{center}
\includegraphics[height=0.67\textheight,width=5in]{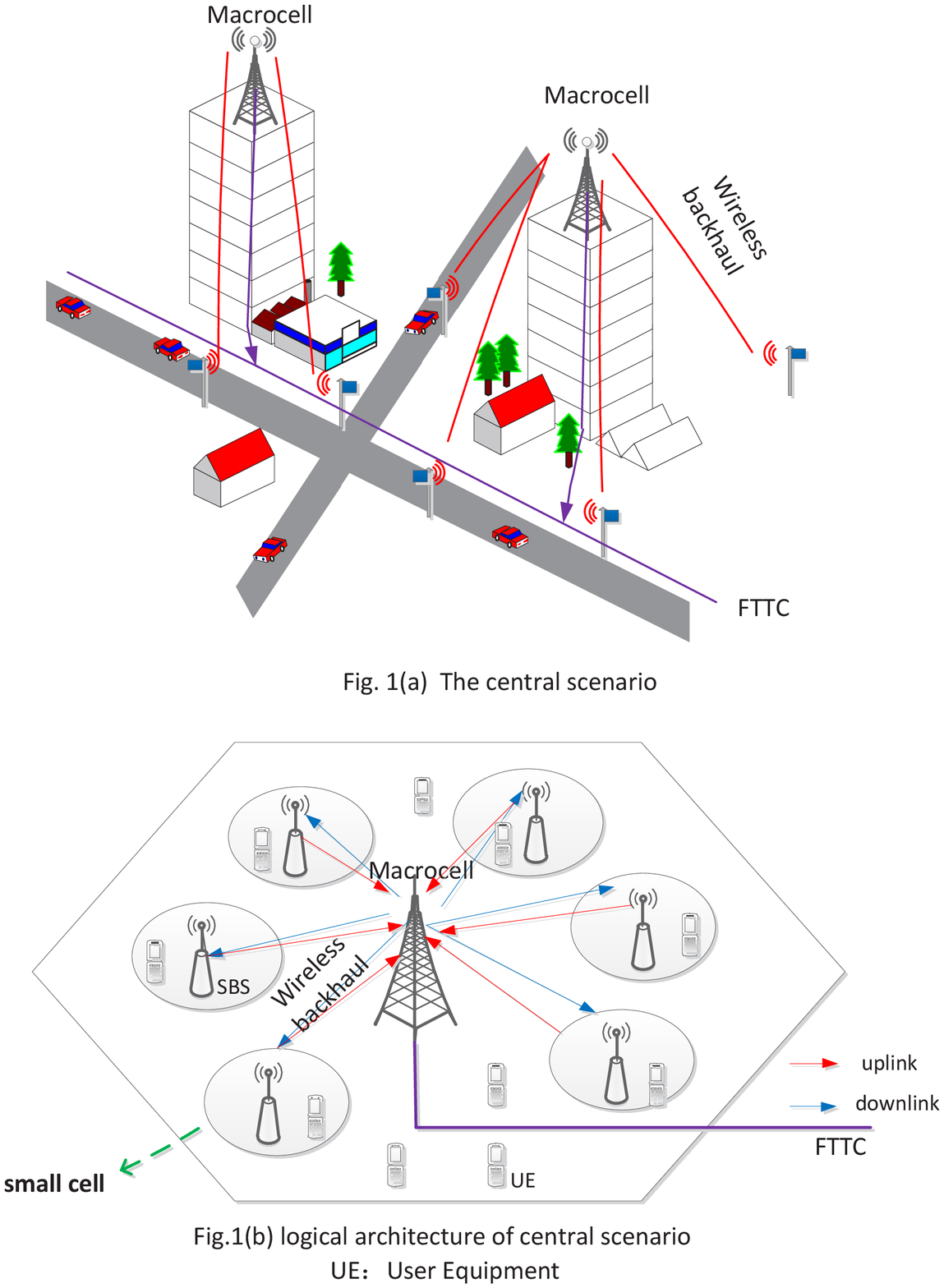}
\caption{The central solution of 5G wireless backhaul networks.}\label{Fig1}
\end{center}
\end{figure}
With massive MIMO and millimeter wave communication technologies, the small cell scenario is an unavoidable solution for the future 5G network. In this paper, two small cell backhaul solutions are presented to evaluate the throughput and energy efficiency of 5G wireless backhaul networks. The first backhaul solution is marked as the central solution in Fig. 1. A macrocell BS (MBS) is assumed to be located in the macrocell center and SBSs are assumed to be uniformly distributed in the macrocell. All SBSs are configured with the same transmission power and coverage. In Fig. 1, the wireless backhaul traffic of small cells is transmitted to the MBS by millimeter wave communication links and then the aggregated backhaul traffic at the MBS is forwarded to the core network by fiber to the cell (FTTC) links. There are two logical interfaces, i.e., S1 and X2 which are used for forwarding backhaul traffic in the central solution. S1 serves as a feeder for user data from the advance gateway to the MBS and the advance gateway is the entrance of the core network. X2 enables mutual information to exchange among small cells. The detailed scenario and the logical architecture are illustrated in Fig. 1(a) and Fig. 1(b).

\begin{figure}
\begin{center}
\includegraphics[height=0.67\textheight,width=5.5in]{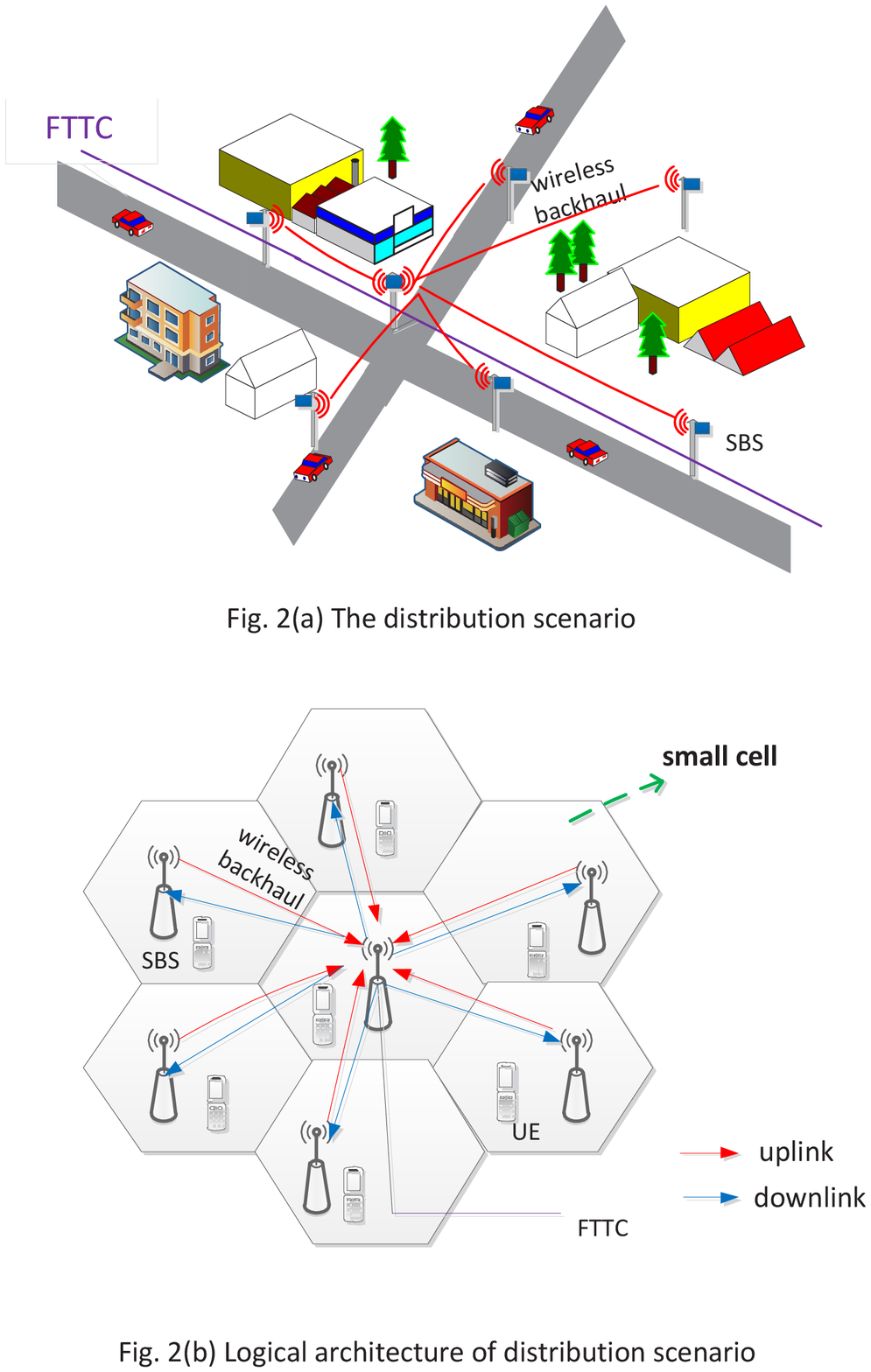}
\caption{The distribution solution of 5G wireless backhaul networks.}\label{Fig2}
\end{center}
\end{figure}

The second backhaul solution is marked as the distribution solution in Fig. 2. Compared with the central solution in Fig. 1, there is no MBS to collect all backhaul traffic from small cells and all backhaul traffic is relayed to a specified SBS. In Fig. 2, all SBSs are assumed to be uniformly distributed in a given area. The backhaul traffic of a SBS is relayed by adjacent SBSs using millimeter wave communications. All backhaul traffic from adjacent SBSs will be cooperatively forwarded to a specified SBS which is connected to the core network by FTTC links. There are two logical interfaces, i.e., S1 and X2 which are used for backhaul traffic in the distribution solution. Functions of S1 and X2 are the same in the central solution and in the distribution solution. The detailed scenario and logical architecture are illustrated in Fig. 2(a) and Fig. 2(b).

\section{Backhaul Traffic Models}

Although the backhaul traffic is comprised of different types of traffic in small cell networks, most of traffic is originated from the user data traffic. However, the overhead backhaul traffic caused by transmission protocols at S1 interfaces and the handover backhaul traffic between adjacent small cells are important parts in the backhaul traffic. In addition, the wireless traffic used for management and synchronization is ignored in this paper since these traffic are obviously less than other traffic in small cell networks. Considering ideal wireless backhaul links between small cells and the
MBS or the specified SBS, the user data traffic is assumed to be only related with the bandwidth and the average spectrum efficiency in every cell. Without loss of generality,
all small cells are assumed to have the same bandwidth and the same average spectrum efficiency. In this case, the backhaul throughput of a small cell is simplified as the production of
the bandwidth and the average spectrum efficiency in the small cell \cite{Robson11,Robson12}. According to results in \cite {Jungnickel13}, the 10\% overhead backhaul traffic is assumed to be generated at S1
interfaces and the 4\% handover backhaul traffic is assumed to be generated at X2 interfaces in small cell networks.

\subsection{Backhaul Traffic Model in Central Solutions}
The backhaul traffic of the central solution includes the uplink and downlink traffic in the macrocell and the small cells. The uplink throughput of a small cell is denoted as $TH_{small - up}^{centra} = 0.04 \cdot B_{sc}^{centra} \cdot S_{sc}^{centra}$, where $B_{sc}^{centra}$ is the bandwidth of a small cell and $S_{sc}^{centra}$ is the average spectrum efficiency of a small cell \cite{Robson12}. The downlink throughput of a small cell is calculated by $TH_{small - down}^{centra} = (1 + 0.1 + 0.04) \cdot B_{sc}^{centra} \cdot S_{sc}^{centra}$, which is transmitted through the S1 interface of backhaul networks. The uplink throughput of a macrocell is denoted as $TH_{macro - up}^{centra} = 0.04 \cdot B_{mc}^{centra} \cdot S_{mc}^{centra}$, where $B_{mc}^{centra}$ is the macrocell bandwidth and $S_{mc}^{centra}$ is the average spectrum efficiency of a macrocell \cite{Robson11}. The downlink throughput of a macrocell is calculated by  $TH_{macro - {\text{down}}}^{centra} = (1 + 0.1 + 0.04) \cdot B_{mc}^{centra} \cdot S_{mc}^{centra}$, which is transmitted through the S1 interface of backhaul networks. Assume that the backhaul traffic is balanced in every small cell. The total number of small cells in a macrocell is configured as $N$. The total uplink backhaul throughput of the  central solution is calculated as $TH_{sum - up}^{centra} = N \cdot TH_{small - up}^{centra} + TH_{macro - up}^{centra}$ and the total downlink backhaul throughput of the central solution is calculated as $TH_{sum - down}^{centra} = N \cdot TH_{small - down}^{centra} + TH_{macro - down}^{centra}$. As a consequence, the total backhaul throughput of the central solution is summed as
$TH_{sum}^{centra} = TH_{sum - up}^{centra} + TH_{sum - down}^{centra}$.

\subsection{Backhaul Traffic Model in Distribution Solutions}
In the distribution solution, adjacent small cells cooperatively forward the backhaul traffic to a specified SBS. Therefore, not only the channel information but also the user data are shared in adjacent cooperative SBSs. Without loss of generality, adjacent cooperative small cells are structured in a cooperative cluster and the number of adjacent small cells in a cluster is assumed as $K$. Without including the specified SBS which collects all backhaul traffic from adjacent small cells, the spectrum efficiency of cooperative cluster is denoted as $S_{sc}^{Comp} = (K - 1)S_{sc}^{dist}$, where $S_{sc}^{dist}$ is the spectrum efficiency of the small cell in the cooperative cluster. Considering the cooperative overhead in the cooperative cluster, the uplink backhaul throughput of a cooperative small cell is denoted as $TH_{small - up}^{dist} = 1.14 \cdot B_{sc}^{dist} \cdot S_{sc}^{dist}$, where $B_{sc}^{dist}$ is the bandwidth of the small cell \cite{Jungnickel13}. The downlink backhaul throughput of a cooperative small cell is denoted as $TH_{small - down}^{dist} = 1.14 \cdot B_{sc}^{dist} \cdot (S_{sc}^{dist} + S_{sc}^{Comp})$. Therefore, the total backhaul throughput of a distribution solution is summed as $TH_{sum}^{dist} = K \cdot (TH_{small - up}^{dist} + TH_{small - down}^{dist})$.

\section{Energy Efficiency of 5G wireless Backhaul Networks}

The energy consumption of cellular networks should include the operating energy and the embodied energy \cite{Humar11}. In this paper, the operating energy is defined as ${E_{OP}} = {P_{OP}} \cdot {T_{lifetime}}$ , where ${P_{OP}}$  is the BS operating power and  ${T_{lifetime}}$ is the BS lifetime. Without loss of generality, the BS operating power is assumed as the linear function of the BS transmission power ${P_{TX}}$ , which is expressed as ${P_{OP}} = a \cdot {P_{TX}} + b,{\text{ }}a > 0{\text{ and }}b > 0$. In general, the transmission power depends on the radius of coverage and the signal propagation fading.

\begin{table}[H]
\centering
\normalsize{
\caption{ Parameters of wireless backhaul networks.}\label{table1}
{\begin{tabular}{ccccc}
\hline
\rowcolor{mygray}
Wireless backhaul frequencies   &  5.8 GHz       &  28 GHz        &  60 GHz  \\
\hline
${a_{macro}}$                   &  21.45         &  21.45         &  21.45  \\
\rowcolor{mygray}
${b_{macro}}$	                &  354 W       &  354 W      &  354 W\\

$P_{TX}^{macro}$
(coverage radius is 500m)         &  10 W	         &  233 W          &  1070 W\\
\rowcolor{mygray}
$P_{OP}^{macro}$
(coverage radius is 500m)          &  568 W 	     &  5352 W      &  23305 W	 \\

$E_{EMinit}^{macro}$	        &  75 GJ          &	75 GJ          &  75 GJ      \\
\rowcolor{mygray}
$E_{EMma\operatorname{int} }^{macro}$	  &  10 GJ         &  10 GJ    &  10 GJ     \\

$T_{lifetime}^{macro}$          &  10 years       &  10 years       &  10 years    \\
\rowcolor{mygray}
${a_{small}}$	                &  7.84          &  7.84          &  7.84\\

${b_{small}}$	                &  71 W	     &  71 W	      &  71 W  \\
\rowcolor{mygray}
$P_{TX}^{small}$
(coverage radius is 50m)	        &  6.3 mW	     &  147 mW         &  675 mW  \\

$P_{OP}^{small}$
(coverage radius is 50m)	        &  71 W	     &  72 W	      &  76 W    \\
\rowcolor{mygray}
\makecell{$E_{EMinit}^{small} + E_{EMma\operatorname{int} }^{small}$\\
( percentage in total energy consumption ) }              &  20\%         &  20\%         &  20\%  \\

$T_{lifetime}^{small}$          &5 years          & 5 years         &  5 years \\
\rowcolor{mygray}
\hline
\end{tabular}\label{table1}
}
\vspace{0.8cm}\\

}
\end{table}

To simplify the model derivation, the MBS transmission power is normalized as ${P_0} = 40$ Watt (W) with the coverage radius ${r_0} = 1$ Kilometer (Km) \cite{Bojic13}. Similarly, the BS transmission power with the coverage radius $r$ is denoted as ${P_{TX}} = {P_0} \cdot {(r/{r_0})^\alpha }$  , where  $\alpha $ is the path loss coefficient. Furthermore, the BS operating power with the coverage radius $r$ is expressed as  ${P_{OP}} = a \cdot {P_0} \cdot {(r/{r_0})^\alpha } + b$. The BS embodied energy includes the initial energy and the maintenance energy, which is calculated by ${E_{EM}} = {E_{EMinit}} + {E_{EMma\operatorname{int} }}$.
In the central solution, the system energy consumption is expressed by\[\begin{split}
 E_{system}^{centra}& = E_{EM}^{macro} + E_{OP}^{macro} + N \cdot (E_{EM}^{small} + E_{OP}^{small}) \\
 &= E_{EMinit}^{macro} + E_{EMma\operatorname{int} }^{macro} + P_{OP}^{macro} \cdot T_{lifetime}^{macro} \\
 &+ N \cdot (E_{EMinit}^{small} + E_{EMma\operatorname{int} }^{small} + P_{OP}^{small} \cdot T_{lifetime}^{small}) \\
\end{split}. \tag{1}\]
Considering the wireless backhaul throughput in the central solution, the energy efficiency of the central solution is defined as ${\eta _{centra}} = {{TH_{sum}^{centra}} \mathord{\left/
 {\vphantom {{TH_{sum}^{centra}} {E_{system}^{centra}}}} \right.
 \kern-\nulldelimiterspace} {E_{system}^{centra}}}$ .

In the distribution solution, the system energy consumption is expressed by \[\begin{split}
  E_{system}^{dist}& = K \cdot (E_{EM}^{small} + E_{OP}^{small}) \\
   &= K \cdot (E_{EMinit}^{small} + E_{EMma\operatorname{int} }^{small} + P_{OP}^{small} \cdot T_{lifetime}^{small}) \\
\end{split}. \tag{2}\]
Considering the wireless backhaul throughput in the distribution solution, the energy efficiency of the distribution solution is defined as ${\eta _{dist}} = {{TH_{sum}^{dist}} \mathord{\left/
 {\vphantom {{TH_{sum}^{dist}} {E_{system}^{dist}}}} \right.
 \kern-\nulldelimiterspace} {E_{system}^{dist}}}$.

To analyze the energy efficiency of 5G wireless backhaul networks in two backhaul solutions, default parameters are configured as follows: the radius of the small cell is 50 meter (m), the macrocell radius is 500 m, the bandwidth of the macrocell and the small cell is 100 Mbps, the average spectrum efficiency of the macrocell is 5 bit/s/Hz \cite{Jungnickel13}, and the path loss coefficient is 3.2 for the urban environment \cite{Misra14}. In macrocells, parameters of BS operating power are configured as $a = 21.45$ and $b = 354.44$ W, respectively. In small cells, parameters of BS operating power are configured as $a = 7.84$ and $b = 71.50$ W, respectively. The lifetime of MBS and SBS are assumed as 10 and 5 years, respectively \cite{Khirallah11}. Other parameters are listed in Table I.

\begin{figure}[H]
\begin{center}
\includegraphics[width=5.5in]{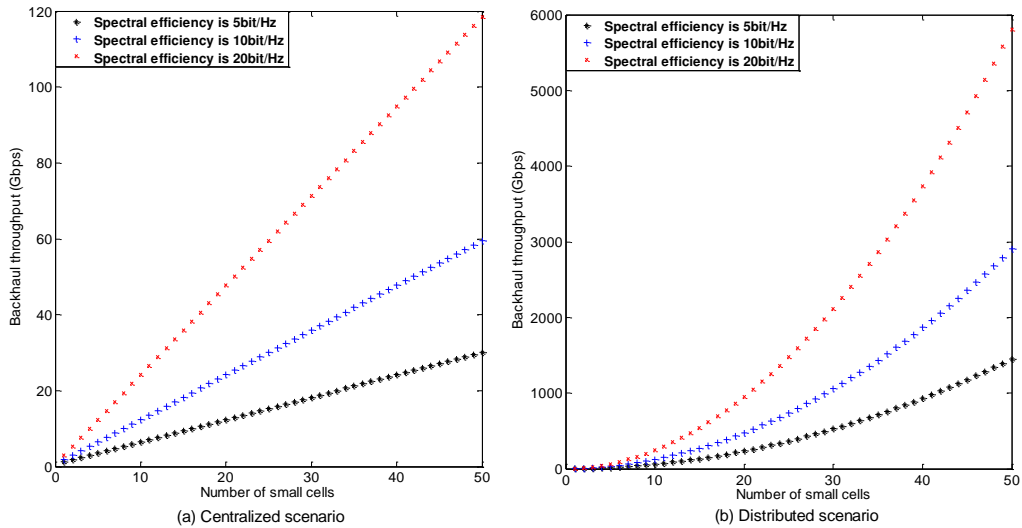}
\caption{Throughput of wireless backhaul networks with respect to the number of small cells considering different average spectrum efficiencies.}\label{Fig3}
\end{center}
\end{figure}

First, the throughput of wireless backhaul networks with respect to the number of small cells considering different average spectrum efficiency is compared in Fig. 3. In Fig. 3(a), the backhaul throughput linearly increases with the increase of small cell numbers in the central solution. In Fig. 3(b), the backhaul throughput exponentially increases with the increase of small cell numbers in the distribution solution. The exponentially increasing feature is caused by the sharing cooperative traffic among small cells in the distribution solution. When the number of small cells is fixed, the backhaul throughput increases with the increase of average spectrum efficiency in small cells.

Second, the energy efficiency of wireless backhaul networks with respect to the number of small cells considering different frequency bands is illustrated in Fig. 4. In Fig. 4(a), the energy efficiency of wireless backhaul networks logarithmically increases with the increase of the number of small cells in the central solution. In Fig. 4(b), the energy efficiency of wireless backhaul networks linearly increases with the increase of the number of small cells in the distribution solution. When the number of small cells is fixed, the energy efficiency of wireless backhaul networks decreases with the increase of frequency bands. However, there exists obviously energy efficiency gaps for 5.8 GHz, 28 GHz and 60 GHz frequency bands in the central solution.

\begin{figure}[H]
\begin{center}
\includegraphics[width=5.5in]{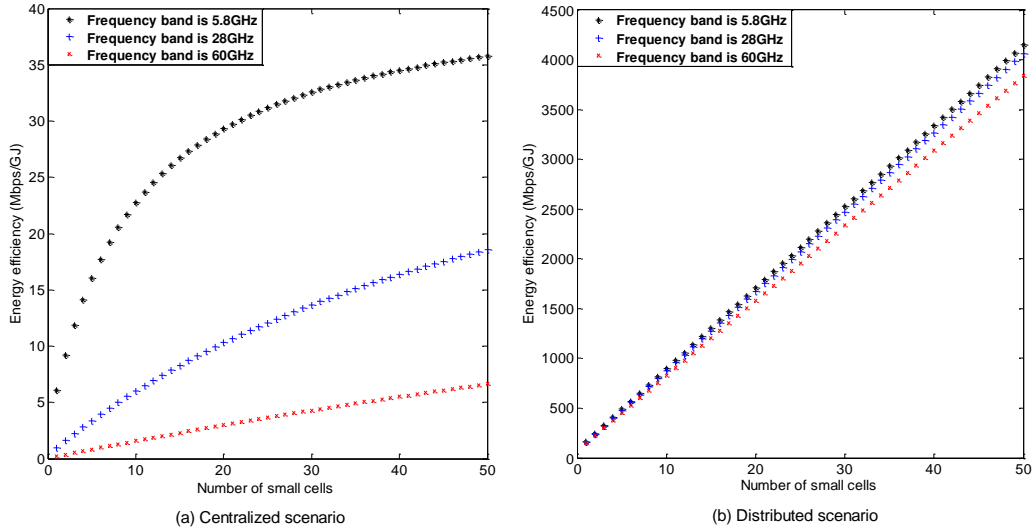}
\caption{Energy efficiency of wireless backhaul networks with respect to the number of small cells considering different frequency bands.}\label{Fig4}
\end{center}
\end{figure}

Finally, the energy efficiency of wireless backhaul networks with respect to the path loss coefficient considering different small cell radii is shown in Fig. 5. When the radius of small cells is less than or equal to 50 m, the energy efficiency of wireless backhaul networks increases with the increase of the path loss coefficient. When the radius of small cells is larger than 50 m, the energy efficiency of wireless backhaul networks decreases with the increase of the path loss coefficient. The reason for this result is that, based on the Shannon capacity theory, the increase of path loss coefficients have a little attenuation effect on the wireless capacity when the radius of small cells is less than or equal to 50 m. In contrast, the increasing path loss coefficients have obviously an attenuation effect on the wireless capacity when the radius of small cells is larger than 50 m. When the system energy consumption is fixed, the energy efficiency is proportional to the wireless capacity in wireless backhaul networks. Compared with central and distribution solutions in Fig. 5(a) and Fig. 5(b), the energy efficiency of the central solution is obviously less than the energy efficiency of the distribution solution under the same radius of small cells and the path loss coefficient.

\begin{figure}[H]
\begin{center}
\includegraphics[width=5.5in]{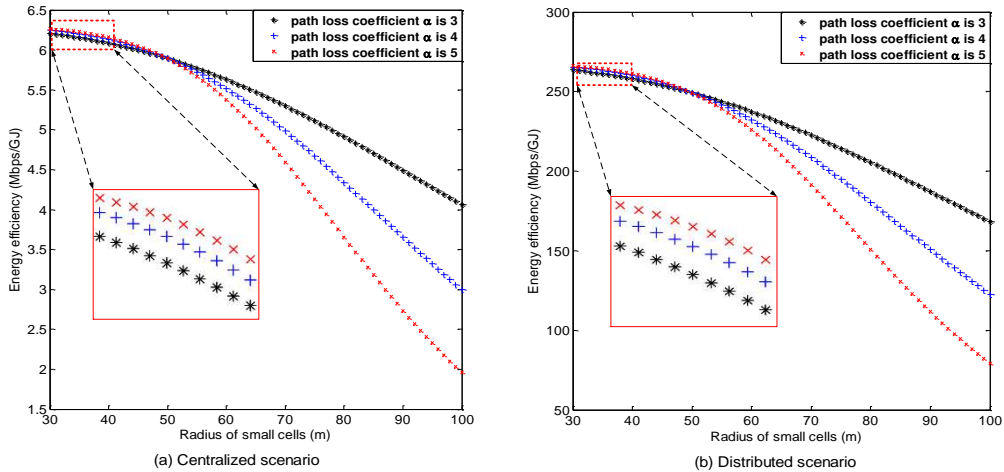}
\caption{Energy efficiency of wireless backhaul networks with respect to the path loss coefficient considering different small cell radii.}\label{Fig5}
\end{center}
\end{figure}

\section{Future Challenges}
In existing studies, the small cell network is presented to satisfy the high capacity requirement for hot areas in cellular networks. Therefore, the large wireless traffic is just transmitted in limited hot areas. In this case, a little burst backhaul traffic can be directly returned back to the core network through the conventional cellular network architecture. With the massive MIMO and millimeter wave communication technologies emerging into 5G networks, the cell size of 5G networks has to become smaller. Furthermore, the large wireless traffic is transmitted in 5G networks. Moreover, the 5G network will become an ultra dense cell network with small cells. As a consequence, it is a great challenge for future 5G wireless backhaul networks to forward massive wireless traffic to core networks in a low cost and high energy efficiency manner. Some potential challenges are presented in the following context.

The first challenge is how to design a new backhaul network architecture and protocols for ultra dense cell deployment scenarios. As we discussed in the last paragraph, small cells and ultra dense deployment will become the main features in future 5G networks. In this case, the number of small cells will obviously increase in the unit area. As a consequence, the corresponding backhaul traffic will increase exponentially at the gateway if the conventional centralized control model is adopted in the 5G backhaul network architecture. The massive backhaul traffic not only makes a congestion but also collapses the backhaul network. It looks that the distributed control model has to be adopted in the 5G backhaul network architecture. However, it brings another challenge whether existing network protocols can support the massive backhaul traffic by wireless links.

For high speed users, how to overcome the effect of frequently handover in small cells is the second challenge. To solve this challenge, the concept of cooperative small cell group is presented to support the high speed user handover among small cells. In this case, multiple small cells must cooperatively transmit traffic to a high speed user. When the high speed user departs a small cell, other cooperative small cells still cover its track and consecutively transmit the high traffic to the high speed user. Moreover, the new small cell is added into the cooperative small cell group based on the high speed user track. However, there are many issues needed to be solved for realizing this idea, such as how to organize a dynamic cooperative cell group and how to decrease the overhead of sharing data in the cooperative small cell group.

Even the massive wireless backhaul traffic can be transmitted back to the core network with a specified QoS, it is a key challenge to deploy it in a high energy efficiency way. Some studies indicate the low power BS with densely deployment will decrease the energy consumption in 5G networks. However, based on our analytical results, different architectures of backhaul networks have different energy efficiency models. For example, in the central solution, the energy efficiency of wireless backhaul networks reach a saturation limit when the density of small cell is larger than a specified threshold. How to optimize the energy efficiency of wireless backhaul networks is very important for the future 5G networks. Some potential solutions are presented to face this energy efficiency challenge, such as hybrid backhaul networks including wireless and fiber links is a valuable solution. Moreover, the new sleeping model in small cells and adaptive power control of SBSs are effective approaches for saving energy in 5G wireless backhaul networks. But more studies should be carried out in the near future.

To overcome the above challenges in 5G wireless backhaul networks, some potential solutions are summarized to solve these issues:
\begin{enumerate}
\item   Distribution cell architecture and protocols can be explored to forward wireless backhaul traffic in the ultra dense small cell networks;
\item Millimeter wave communications are recommended to transmit the massive backhaul traffic in 5G wireless backhaul networks;
\item Cooperative small cell groups should be investigated to solve handover issues in small cell networks;
\item High energy efficiency transmission technologies should be developed to guarantee 5G wireless backhaul networks deploying in a low energy consumption way.
\end{enumerate}

\section{Conclusion}
5G networks are expected to satisfy rapid wireless traffic growth. Massive MIMO, millimeter wave communications and small cell technologies are presented to achieve Gigabit transmission rate in 5G networks. In this paper, we study how to promote 5G wireless backhaul networks in high throughput and low energy consumption ways. Two typical small cell scenarios are configured to analyze the wireless backhaul traffic in future 5G networks. Furthermore, the energy efficiency of wireless backhaul networks is compared by two typical small cell scenarios. Numerical results imply that the distribution solution has higher energy efficiency than the central solution in 5G wireless backhual networks. However, a veritable challenge would indeed emerge if the new distribution network architecture is adopted in the future 5G wireless backhaul networks.

\begin{IEEEbiographynophoto}{Xiaohu Ge} [M'09-SM'11] is currently a Professor with the Department of Electronics and Information Engineering at Huazhong University of Science and Technology (HUST), China. He received his Ph.D. degree in Communication and Information Engineering from HUST in 2003. From January 2013, he was granted as a Huazhong scholarship professor. He serves as an Associate Editor for the IEEE ACCESS, Wireless Communications and Mobile Computing Journal, etc.. \end{IEEEbiographynophoto}

\begin{IEEEbiographynophoto}{Hui Cheng} received the Bachelor degrees in information engineering from Wuhan University of Technology, China, in 2013. She is currently working toward the Master degree in communication and information systems at Huazhong University of Science and Technology, Wuhan, China. Her research interests are in the fields of mobile backhaul traffic and user mobility models for small cell networks. \end{IEEEbiographynophoto}

\begin{IEEEbiographynophoto}{Mohsen Guizani} [S'85-M'89-SM'99-F'09] is currently a Professor and the Associate Vice President for Graduate Studies at Qatar University, Qatar. He received his B.S. (with distinction) and M.S. degrees in Electrical Engineering; M.S. and Ph.D. degrees in Computer Engineering in 1984, 1986, 1987, and 1990, respectively, from Syracuse University, Syracuse, New York. His research interests include Computer Networks, Wireless Communications and Mobile Computing, and Optical Networking. He currently serves on the editorial boards of six technical Journals and the Founder and EIC of “Wireless Communications and Mobile Computing” Journal published by John Wiley (http://www.interscience.wiley.com/jpages/1530-8669/). He is an IEEE Fellow and a Senior member of ACM. \end{IEEEbiographynophoto}

\begin{IEEEbiographynophoto}{Tao Han} [M'13] (hantao@hust.edu.cn) received the Ph.D. degree in communication and information engineering from Huazhong University of Science and Technology (HUST), Wuhan, China in December, 2001. He is currently an Associate Professor with the Department of Electronics and Information Engineering, HUST. His research interests include Wireless Communications, Multimedia Communications, and Computer Networks.\end{IEEEbiographynophoto}

\end{document}